# Toward Photon-Induced Near-Field Electron Tomography

Tamir Shpiro[1], Ron Ruimy[1], Qinghui Yan[1], Tomer Bucher[1], Avner Shultzman[1], Hanan Herzig Sheinfux[2], and Ido Kaminer[1*]

1. Solid State Institute, Technion-Israel Institute of Technology, Haifa 32000, Israel
2. Department of Physics, Bar-Ilan University, Ramat Gan 5290002, Israel

*kaminer@technion.ac.il

New techniques for imaging electromagnetic near-fields in nanostructures drive advancements in nanotechnology, optoelectronics, materials science, and biochemistry. Most existing techniques probe near-fields along surfaces, lacking the ability to extract near-fields confined within the structure. Notable exceptions use free electrons to traverse through nanostructures, integrating the field along their trajectories, extracting 2D near-field projections rather than the complete field. Here, drawing inspiration from computed tomography (CT), we present a tomography concept providing full 3D reconstruction of vectorial time-harmonic near-fields. We develop a Radon-like algorithm incorporating the electron wave-nature and the time dependency of its interaction with vector fields. To show the prospects of electron near-field tomography, we propose and analyze its ability to resolve the sub-wavelength zigzag profile of highly confined hyperbolic polaritons and to reconstruct 3D phase singularities in a chiral near-field – raising exciting goals for next-generation experiments in ultrafast transmission electron microscopes.

Introduction

Studies of electromagnetic near-fields in illuminated nanostructures stand at the core of light-matter interactions, with important applications in photonics, material science, nanotechnology, and the life sciences[1]. The full visualization of near-fields in three dimensions

(3D) is particularly intriguing considering the fast-paced growth of nanoscale 3D printing technologies and other nanofabrication processes[2]; these create photonic structures with intricate electromagnetic responses, from chiral[3] or multilayered metasurfaces[4] and heterostructures of 2D materials[5] to silicon photonic waveguides[6] and 3D photonic crystals[7]. The gradual increase in the complexity of photonic structures calls for new methods for the full visualization of their near-fields, especially of the near-fields buried *inside* them. This challenge is most pertinent for complex nanomaterials such as core-shell nanoparticles[8], hybrid multilayered van der Waals materials[9], quantum-dot superlattices[10], and integrated waveguides[6], where the internal field governs their light-matter interactions.

In this study, we will use the term 'near-field' to refer to localized electromagnetic fields that include both the field near the surface of nanostructures as well as fields that penetrate into these structures. This usage is adopted to encompass the full range of electromagnetic phenomena observed in nanostructures, aligning with the broader definitions occasionally employed in our field[11,12].

The two most prolific techniques for near-field imaging are photo-emission electron microscopy (PEEM)[13] and scattering-type scanning near-field optical microscopy (sSNOM)[14], both imaging near-fields along surfaces with sub-wavelength spatial resolution. These techniques are limited to the part of the near-field *outside* the structures, with progress enabling to resolve both phase, amplitude, and polarization[15–21]. Pioneering works demonstrated 3D reconstruction of near-fields, but could only access the fields outside the structure[22], leaving the field confined *inside* the structure unexplored. Exceptions that could access the near-fields inside nanostructures relied on nonlinear light generated by mixing optical illumination with the near-fields[23–26], enabling to extract their phase and polarization information. However, these techniques necessitate the material to have specific optical nonlinearities.

A fast-evolving technique called photon-induced near-field electron microscopy (PINEM)[11,27–30] can access the near-fields confined inside nanostructures with high spatial resolution[12]. PINEM relies on measuring the energy change of electrons undergoing inelastic scattering off the near-fields inside and around illuminated nanostructures (Fig. 1a). Electrons of tens-to-hundreds of keV can penetrate through samples of up to a few hundred nanometers while still maintaining their coherent interaction. Such penetration depths enable PINEM to access the near-fields inside the bulk. Developments inspired by Ramsey (or homodyne) electron interferometry[31–37] pre-modulate the electrons to enable electron imaging of both the amplitude and phase information of the near-fields at every point (Fig. 1b)[38–40]. Even with these recent developments, PINEM so far only extracted field projections, integrating the field polarization component parallel to the electron trajectory. Not only do the other polarization components not participate, but the integration also folds any field information existing along the electron trajectory, leaving the full 3D field profile and its vectorial nature inaccessible.

In the wider field of transmission electron microscopy (TEM)[41], tomography is used to reconstruct the 3D profiles of scalar potentials of thin phase objects, becoming the frontier of TEM in the life sciences, famously in cryogenic electron tomography[42,43]. Electron tomography relies on sequences of images at varying rotations, each providing a different projection of the sample, which are then combined for a 3D reconstruction[44]. The reconstruction algorithms span from conventional computed tomography (CT) algorithms as in X-ray imaging[45], to advanced usages combining it with ptychography[46,47] and machine learning to improve the accuracy and speed of image processing, particularly for handling large datasets and reducing artifacts[48].

Variants of electron tomography also reconstruct 3D electromagnetic vector fields using multiple projections of Lorentz TEM[49,50] and reconstruct the 3D local density of photonic states of surface plasmons[51–53] using projections of cathodoluminescence[54] or electron energy-loss spectroscopy[55,56]. All of these diverse entities are still stationary in time, and thus suitable for conventional CT algorithms[45] and their state-of-the-art extensions.

However, when considering the challenge of tomography of electromagnetic near-fields, previous approaches cannot apply due to the time-dependent oscillating nature of the fields: the electron mixes different field values, as the phase of the optical near-fields completes multiple cycles during the transmission time of the electron. Moreover, the quantum wave nature of the electron imposes an intrinsic uncertainty in its arrival time, which is longer than the field cycle, causing each electron to interact coherently with multiple field values, interfering them together. Additional statistical uncertainty in the electron arrival time prevents phase sensitivity, adding further complexity. These limitations call for a new concept of electron tomography that can account for both the time-dependent vectorial nature of near-fields and for the quantum wave nature of the electrons.

Here we present a novel approach for 3D reconstruction of time-harmonic vector near-fields, denoted as photon-induced near-field electron tomography (PINET) (Fig. 1c). The concept is based on a specialized CT algorithm tailored for the time- and direction- dependence of the inelastic interaction between free-electrons and the electromagnetic fields (Fig. 1d), accounting for their vectorial nature. We analyze the performance and limitations of this tomographic concept in 2D (Figs. 2, 4-5) and 3D (Fig. 3).

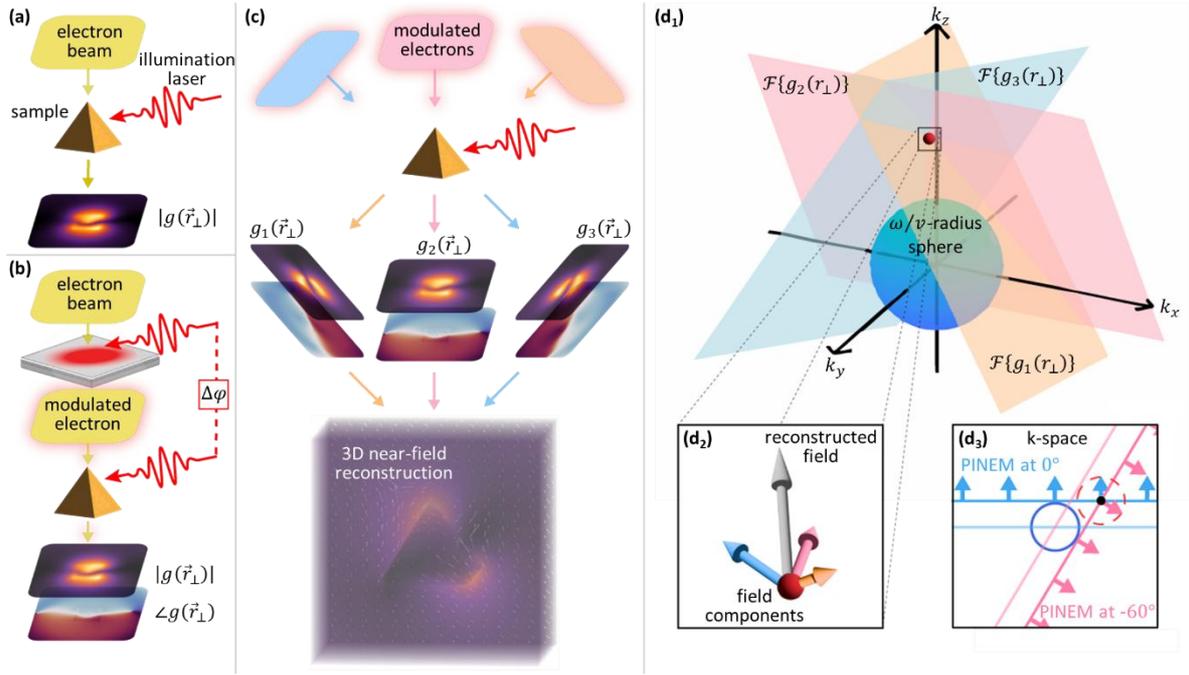

**Fig.1 Photon-induced near-field electron tomography (PINET): concept and algorithm.**
**(a)** Conventional PINEM: The electron integrates the near-field along its trajectory, providing a 2D projection. **(b)** Interferometric PINEM: Scanning over the phase-delay relative to a phase-locked reference field, providing a complex-valued 2D projection[38–40]. **(c)** PINET: Measuring complex-valued 2D projections from multiple angles to reconstruct the entire vectorial complex-valued 3D near-field. **(d)** Geometric scheme illustrating the algorithm idea: We reconstruct the 3D near-field in Fourier space. Each 2D projection by interferometric PINEM corresponds to a plane in k-space tangent to a sphere of radius $\omega/v$ centered at the origin. The electron interaction extracts the inverse Fourier transform of the field polarized normal to this plane, at every point along the plane, denoted by $g(\vec{r}_\perp)$. We record the 2D Fourier transform $\mathcal{F}\{g(\vec{r}_\perp)\}$ of each such plane, together finding the field values in the entire Fourier space outside the sphere of radius $\omega/v$. Fourier values inside this sphere are inaccessible using PINET. **(d2)** To reconstruct the full 3D vectorial field at each point, we combine information from three planes intersecting at that point. **(d3)** 2D slice of the formers, comparing conventional projections (bright lines, intersects at the origin) with PINEM projections (tangent to $\omega/v$-circle, with small arrows).

Building on this tomography concept, we then propose and simulate experiments that can use it in ultrafast transmission electron microscopes: **Tomography of optical singularities** in 3D chiral near-fields hints to an intriguing general conjecture relating topological singularities in 3D near-fields[57,58] to phase singularities in their 2D projections (Fig. 3). **Tomography of hyperbolic polaritons** in 2D materials reveals their famous sub-wavelength zigzag profile that was predicted in theory works[59] and indirectly inferred[60–62] but never directly observed (Fig. 4). **Tomography of near-fields in a complex nanostructure** shows the influence of limiting experimental conditions and allows us to test algorithmic enhancements (Fig. 5). Through the analysis of these eminent near-field phenomena, we discuss the remaining challenges for realizing PINET in current ultrafast electron microscopes.

**Theory of photon-induced near-field electron tomography (PINET)**

The theory of PINET relies on the interaction between paraxial free electrons (with velocity $\vec{v}$) and time-harmonic (i.e., approximately monochromatic) $\vec{E}(\vec{r})e^{-i\omega t}$ optical near-field. The resulting electron energy spectrum at each transverse point $\vec{r}_\perp$ (along a plane perpendicular to $\vec{v}$) is determined by a single dimensionless parameter (interaction strength) $g$ known from the PINEM theory[27,28]:

$$g(\vec{r}_\perp) \equiv \mathcal{P}_{\vec{v}}\{\vec{E}\}(\vec{r}_\perp) = \frac{e}{\hbar\omega}\int_{-\infty}^{\infty} \vec{v}\cdot\vec{E}(\vec{r}_\perp + \vec{v}t)e^{-i\omega t}dt. \quad (1)$$

This integral yields a phase-matching condition implying that electrons with velocity $\vec{v}$ only interact with Fourier k-components satisfying $\vec{k}\cdot\vec{v} = \omega$. Since the electrons are paraxial, their energy spectrum is only affected by the field polarization parallel to their trajectory ($\hat{v}\cdot\vec{E}$).

The electron energy spectrum is highly nonlinear in $g(\vec{r}_\perp)$. The probability for an energy shift $\Delta E = n\hbar\omega$, with an integer $n$, is $P_n(\vec{r}_\perp) = J_n^2(2|g(\vec{r}_\perp)|)$, where $J_n$ is the Bessel function of order $n$. By pre-modulating the electron using reference fields phase-locked to the sample's field, both the phase and amplitude of the projection at every $\vec{r}_\perp$ can be accessed (Fig. 1b). In practice, it is more convenient to extract energy-filtered images (integrated above a chosen threshold) than the full spectrum at each position. In the case where $g(\vec{r}_\perp) \ll 1$, the energy-filtered signal becomes approximately linear in $g(\vec{r}_\perp)$, resulting in direct access to the field's phase[38,40]. This is not applicable in most cases where the signal is non-linear in $g(\vec{r}_\perp)$. Accounting for this nonlinearity, algorithmically-based methods[37] enable extracting both amplified amplitude and phase[39] of the projection, for all regimes.

Notice that our projection operator $\mathcal{P}_{\vec{v}}$ (Eq. 1) has two main differences compared to the conventional projection assumed in CT algorithms: [1] The projected quantity is vectorial, and the projected component depends on the angle of projection ($\hat{v} \cdot \vec{E}$). [2] The projected quantity varies during the electron transmission due to the $e^{-i\omega t}$ factor in the integral. Under the assumption of time-harmonic fields, this projection provides a complex-valued Fourier component of the field. In comparison, ordinary CT provides a real-valued integral (the 0-frequency Fourier component). The integral of ordinary CT is retrieved in the limit of $v \to \infty$, or $v \gg \omega L$, where $L$ is the effective support of the integral. This condition is not relevant for optical fields in nanostructures, but may be possible for lower frequencies.

As noted, the standard CT algorithm is not applicable for near-fields. Nevertheless, the mathematical idea behind it provides the blueprint for developing the PINET algorithm. The conventional CT algorithm is based on a mathematical theorem denoted as "the projection-slice theorem"[45]. The idea is as follows: Assuming a 2D scalar object with density $\rho(x, y)$, and a projection $m(x) = \int \rho(x, y) dy$, Fourier transforming $m(x)$ gives $M(k_x) = P(k_x, 0)$, where $M$ and $P$ are the Fourier transforms of $m$ and $\rho$. The projection $m(x)$ provides a "slice" $P(k_x, 0)$ of the complete $P(k_x, k_y)$. Every rotation in 2D real-space converts to the exact same rotation in 2D k-space (Fourier space). That way, a series of projections from different directions translates in k-space to a series of rotated slices intersecting at the origin, filling the whole k-space (Fig. 1d$_3$). To fill up the full 2D k-space using a finite number of slices, interpolation methods are required, as have been researched extensively in CT[63]. After getting P for every $(k_x, k_y)$, inverse Fourier transform provides the required density distribution $\rho(x, y)$. Improved algorithms have been developed, e.g., assuming extra information regarding the reconstructed object[64].

Taking this idea into our case, the analog of "slice" here is a "directional slice", defined by the operator $\mathcal{S}_{\vec{v}}\{\vec{E}\}(\vec{k}_\perp) = \frac{e}{\hbar\omega}\hat{v} \cdot \vec{E}\left(\vec{k}_\perp + \frac{\omega}{v}\hat{v}\right)$ acting on the vector field $\vec{E}$. The Fourier transform then provides (proof in SI):

$$\mathcal{F}\left\{\mathcal{P}_{\vec{v}}\{\vec{E}\}\right\} = \mathcal{S}_{\vec{v}}\left\{\mathcal{F}\{\vec{E}\}\right\}. \tag{2}$$

Since $g(\vec{r}_\perp)$ is itself a Fourier component of the $\hat{v} \cdot \vec{E}$ field polarization along the $\hat{v}$ direction, its 2D Fourier transform on $\vec{r}_\perp$ provides a 2D k-space map of Fourier components of $\hat{v} \cdot \vec{E}$. The 2D map corresponds to a plane in the 3D k-space of the field, shifted from the origin by a wavevector of size $\omega/v$ and direction of the electron motion relative to the sample (Fig. 1d$_1$). Geometrically, all these planes tangent to the same $\omega/v$-radius sphere centered at the origin (Fig. 1d$_1$). Outside that sphere, each point is contained inside an infinite number of planes tangent to the sphere (or exactly two in the 2D case illustrated in Fig. 1d$_3$). Combining the information from three different tangent planes (or two for the 2D case) passing through the point in k-space, we retrieve the full vectorial information at that point (Fig. 1d$_2$). This approach provides the reconstruction of the full complex and vectorial information in k-space, outside the $\omega/v$-radius sphere. We note that unlike in ordinary CT, the phase information of the projections is crucial to get the full field information.

The missing $\omega/v$-radius sphere in k-space (Figs. 1d$_1$ and 1d$_3$) is an intrinsic limitation of a projection-based tomography of time-harmonic objects. These spatial near-field frequencies cannot be probed by free electrons with velocity $v$ because they do not satisfy phase-matching for any angle of incidence. Mathematically, we see this limitation by a Cauchy–Schwarz inequality: $\|\vec{k} \cdot \vec{v}\| \leq kv < \omega$ for any $\vec{k}$ inside $\omega/v$-radius sphere. This limitation is general, independent of the specific probe particle or the reconstructed time-harmonic object. The reconstructed near-field is equivalent to the full field after a high pass $\omega/v$ filter. However, as will be shown later, this filter conserves all small features (for optical frequencies and semi-relativistic electrons, smaller than a few hundred nanometers), which are the most interesting for confined near-fields.

**Results**

*Demonstration on a 2D nanostructure*

To exemplify PINET, we apply it to reconstruct near-fields in different illuminated nanostructures. We start with near-field scattered from an illuminated gold nano-wire that is long enough to be described effectively by a 2D field. To collect projections, we rotate the nano-wire in coordination with the illumination around its axis, perpendicular to the electron path. Figs. 2$b_2$, 2$c_2$ and 2$d_2$ compare the scattered electric field, the confined electric field (i.e., only spatial frequencies higher than $\omega/c$, or equivalently phase velocity smaller than $c$), and the electron-accessible field (same with $\omega/v$ and $v$ respectively). The respective fields in k-space are displayed in Figs. 2$b_1$, 2$c_1$ and 2$d_1$. Fig. 2e shows the improvement of the reconstruction for better TEM resolution and for more measurement angles. Good results can be seen already with only 7 measured angles. The error approaches zero for infinitesimal resolution. To mitigate high discretization errors caused by sensitivity to high-frequency noise, we apply a smoothing filter to the reconstructed image.

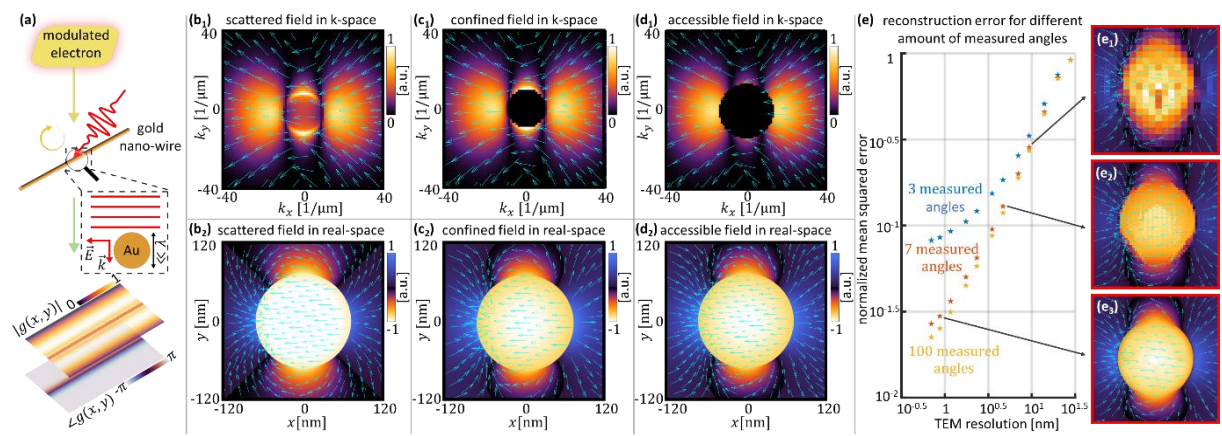

**Fig.2 PINET for an effectively 2D structure: showing the trade-off between number of angles and precision. (a)** PINEM measurements of a long gold nano-wire illuminated by a planewave (polarization and k-vector both perpendicular to the wire axis). The long dimension

of the wire makes this configuration effectively 2D. **(b)** The scattered electric field in k-space and real-space. **(c)** The confined electric field in k-space and real space (effectively, a high-pass-$\omega/c$-filtered field). **(d)** The electric field that can interact with the electron in k-space and real space (effectively, a high-pass-$\omega/v$-filtered field). **(e)** Normalized mean squared error value of the reconstruction as a function of the TEM resolution and the number of measured angles (different colors), showing gradual converging to perfect reconstruction. Both axes are log scaled. All panels assume 200 keV electrons, a 140 nm diameter gold nano-wire, and a 700 nm laser illumination.

*Demonstration on a 3D nanostructure, extracting field singularities*

To demonstrate our approach in 3D, we apply PINET on the near-field of an illuminated gold nano-pyramid. This nanostructure is used as a platform for exploring a more general concept: tomography of an optical field singularity (Fig. 3). Field singularities are a universal concept ubiquitous to many wave systems. Exploration of electromagnetic wave singularities intersects advanced optics, materials science, and wave physics. These singularities play a crucial role in the development of technologies ranging from optical manipulation tools[65] to high-resolution microscopy[66] and optical communications[67]. In 2D systems, various types of singularities have been demonstrated, from polariton vortices[68–71] and Möbius rings[72] to skyrmions[17], carrying a conserved topological charge. The measurement of these singularities in both the far-field[72] and near-field[17,39] most often focused on their 2D nature. Extending the concept to 3D introduces increased complexity. 3D singularities such as optical vortex knots[57] and hopfions[58] represent just a small subset of the intricate topological structures that can occur in light fields. These singularities are harder to measure and analyze.

The illumination of the gold nano-pyramid breaks its symmetry, making the combined system (illumination + pyramid) chiral. The chiral nature of the system enables a pair of 3D field singularities to appear on opposite pyramid faces (Fig. 3a). First evidence of the near-field chirality comes from observing the odd number of phase singularities in perpendicular projections (Fig. 3b$_1$).

Deeper investigation of projections from different directions (Fig. 3b$_2$) hints at a close relation between the 3D field singularities of the near-field and its 2D projections. It is an open question whether a 3D singularity of the field must have 2D singularity in its projections, and whether 2D singularity in a projection necessitates a 3D singularity in the field. In the pyramid field and its projections, both directions of this question are satisfied: there are 3D singularities in the field as well as 2D phase singularities in several projections. However, a general answer requires in-depth research, taking into account the unique projection of PINEM.

Considering experimental implementations, standard TEM holders can have two axes of rotations, one is the symmetry axis of the holder, and the other is perpendicular to it and to the electron beam. The angles, denoted as $\alpha$ and $\beta$ (Fig. 3a), are limited to a specific range, which limits the quality of reconstruction but can be compensated by additional algorithmic improvements as discussed below. Unlike conventional tomography techniques, PINET requires the coordinated rotation of the illumination laser and the nano-pyramid.

Due to the large dimensionality of the data, we present the reconstruction of two of the three field components, through rotations around a single axis (Fig. 3d). Each 2D projection is separated into 1D slices perpendicular to the rotation axis. Then 2D reconstruction is applied to get 2D slices of the field components perpendicular to the rotation axis, creating the displayed 3D field. We get a good fit between the confined field and the reconstructed one, both in 3D view, 2D slices, and along the pyramid faces (Figs. 3c and 3d).

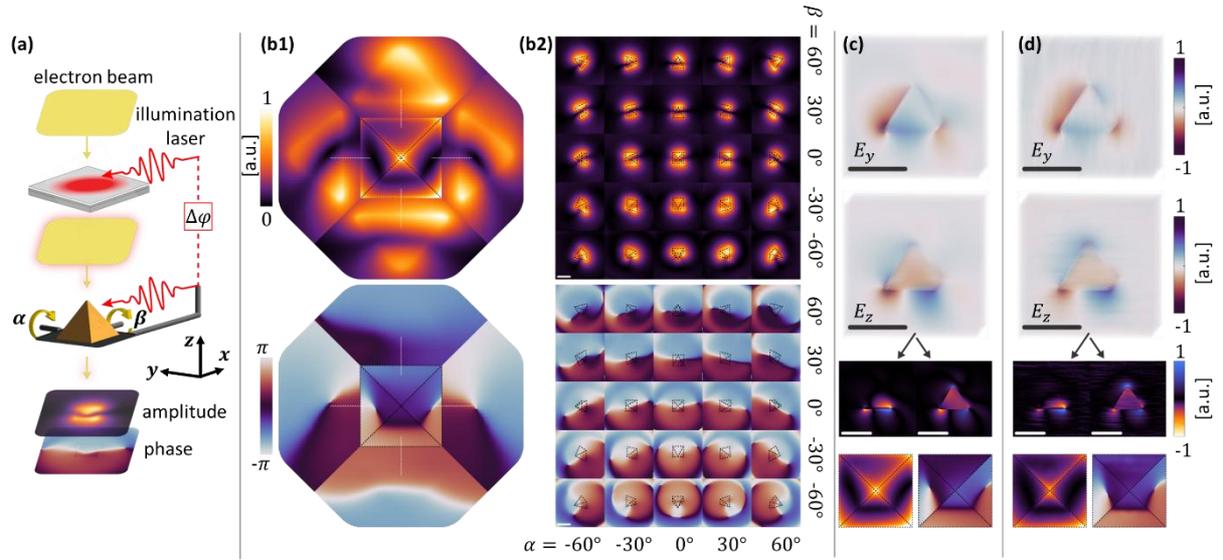

**Fig.3 PINET implemented for a 3D structure: tomography of field singularities. (a)** PINET scheme demonstrated on a gold nano-pyramid, denoting the rotation angles α, β. For each orientation, a scan over multiple phase delays $\Delta\varphi$ provides the complex-valued field projection. The illumination laser is rotated together with the sample. **(b1)** Phase singularities along the pyramid's faces (normal polarization components) and the corresponding electron projections, showing amplitude (top) and phase (bottom). **(b2)** Electron projections for multiple α-β combinations. **(c, d)** Comparing the reconstructed field to the simulated confined field, showing the real parts of the y,z components in a 3D view (top), 2D slices view (center), and field perpendicular to the pyramid's faces (bottom). All panels assume 200 keV electrons and a 700 nm illumination. Scale bars are 200 nm.

*Demonstration on polaritons inside a van der Waals material*

PINET can reconstruct near-fields buried inside the sample. This capability is especially relevant when the fields are guided within the sample and demonstrate intricate propagation dynamics. To demonstrate this point, we analyze the phonon-polariton field guided inside a hexagonal boron nitride (hBN) flake. The dynamics of such polaritons has been extensively explored using near-field microscopy[73–75], and more recently using ultrafast electron microscopy methods[76,71,39] based on PINEM. Due to their hyperbolic dispersion, even in a relatively thin flake there are many phonon-polariton modes. The coherent interference of these multiple modes can manifest as highly localized excitations propagating inside the flake in a "zig-zag" ray-like fashion[59] (Fig. 4a). The ray size is much smaller than the flake width, which can itself be much smaller than the wavelength of the illuminating field. These rays have been researched extensively in recent years, proven by theory and simulations, and indirectly measured[60–62]. However, due to the inability of most current near-field imaging methods to reconstruct the bulk field profile, the "zig-zag" effect has not been directly observed.

We show that PINET (Fig. 4d) directly reconstructs the features of the scattered and guided fields (Fig. 4c), including the full ray path inside, revealing the famous "zig-zag" features. This exemplifies the effect of the intrinsic high-pass filter of PINET, showing that it does not inhibit the reconstruction of the full near-field information, which is mostly contained in features that are extremely small compared to the wavelength.

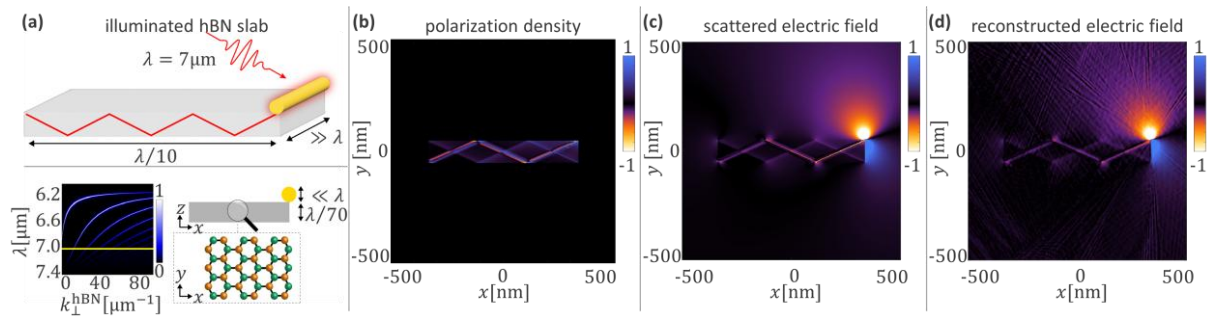

**Fig.4 Reconstruction of intricate near-field propagating inside the bulk. (a)** Hexagonal boron nitride (hBN) flake, illuminated by $\lambda = 7$ μm light coupled from the edge (modeled in 2D as a point-like dipole source), showing a highly confined near-field profile. The excited polaritonic field comprises multiple modes in the hBN's dispersion relation, shown below the schematic of the hBN flake. **(b)** Polariton polarization density in the hBN, showing its highly confined "zig-zag" ray-like propagation. **(c-d)** Simulated scattered and reconstructed electric fields, showing a good fit, including both large and small features. The reconstruction assumes 200 keV electrons.

*Demonstration of enhanced-PINET using a physics-inspired gradient descent optimization*

Finally, we demonstrate how PINET handles intricate near-fields in nanostructures with sharp, sub-wavelength edges, even given a limited angular range. We apply our algorithm on the simulated near-field in a long gold nano-wire (effectively 2D) with the cross-sectional shape of the Technion logo (Fig. 5a). We apply the reconstruction twice. First, with a full range of angles, and second, with a limited range of angles, −45° to 45° (from both sides – up and down). These conditions exemplify typical limitations in TEMs with practical illumination conditions. For the missing angles we first use interpolation between the closest measured angles. In comparison to the full angular range (Fig. 5f), the results in the limited case are poor (Fig. 5g) and call for a better reconstruction algorithm.

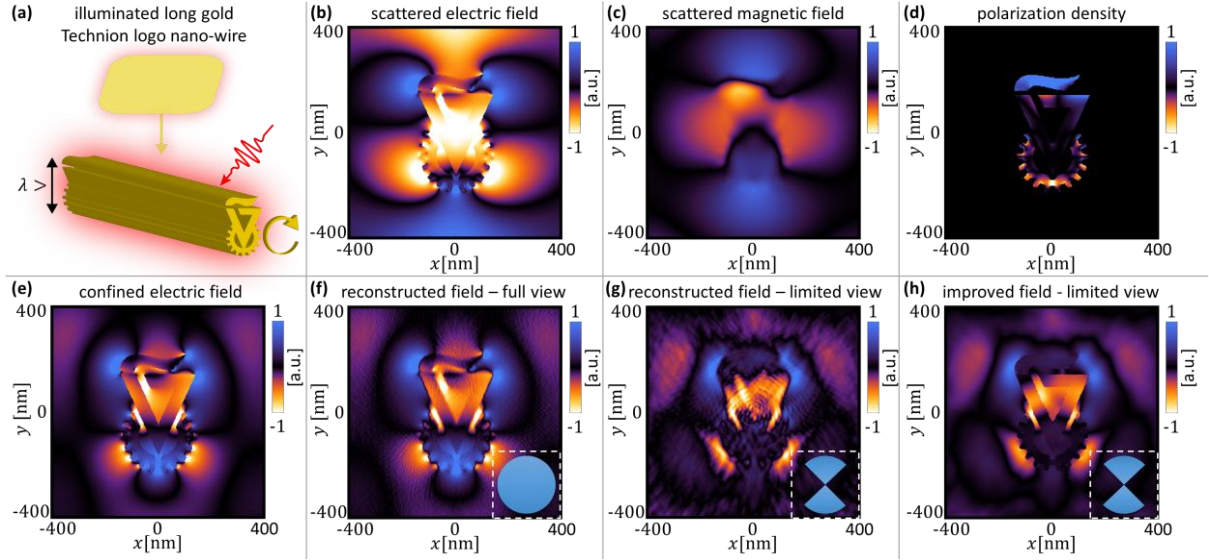

**Fig.5 Physics-inspired gradient descent PINET under limited angular range. (a)** The nanostructure is a long gold nano-wire with a cross-sectional shape of the Technion logo, illuminated by a plane wave (polarization and k-vector both perpendicular to the wire axis). The long dimension of the wire makes this configuration effectively 2D. **(b-d)** The scattered electric and magnetic fields and polarization density, satisfy Maxwell's equations and exhibit piecewise continuity, full continuity, and limited support, respectively. **(e)** The confined electric field. **(f)** Reconstructed electric field based on a full angular range. **(g)** Reconstructed electric field based on a limited angular range ($-45°$ to $45°$ from both sides – up and down). **(h)** improved reconstruction based on the limited angular range, imposing piecewise continuity constraint of the electric field. All panels assume 200 keV electrons, and a 700 nm laser illumination.

We find that a significant improvement in the reconstruction can be achieved by using physical assumptions regarding the near-field. The near-field must satisfy Maxwell's equations, which impose several constraints that the algorithm could rely on. The first constraint is piecewise continuity of the electric field, this arises from the piecewise continuity of the permittivity of the illuminated sample (Fig. 5b). The second constraint is similar to the first, but for the magnetic field. In this case, the constraint becomes full continuity, assuming constant permeability (Fig. 5c). The third constraint is a limited support for the polarization density, which lies only inside the sample (Figs. 4b and 5d). We show an immediate improvement in the results using the first assumption, adding piecewise smoothing (Fig. 5h).

We propose an improved algorithm for solving the tomography challenge based on a different approach – as an optimization problem. We present an iterative algorithm that relies on the physical PINEM model and imposes the constraints detailed above. We define a loss function that presents the measurement information by a fidelity term, the distance between the given measurements and the projections of the current guess. The physical assumptions are represented in the loss function as three more terms representing the constraints detailed above.

To find these terms explicitly, we use Maxwell's equations to isolate relations between the electric field $\vec{E}$, the magnetic field $\vec{H}$, and the polarization density $\vec{P}$:

$$i\omega\mu_0 \vec{H} = \vec{\nabla} \times \vec{E}, \quad \omega^2\mu_0\vec{P} = \vec{\nabla} \times \vec{\nabla} \times \vec{E} - \left(\frac{\omega}{c}\right)^2 \vec{E}. \tag{3}$$

Then, one way to write the loss function is as follows:

$$\text{Loss}(\vec{E}) = c_0 L_{\text{fidelity}}(\vec{E}) + c_1 L_E(\vec{E}) + c_2 L_H(\vec{E}) + c_3 L_P(\vec{E}). \tag{4}$$

The first term, called the fidelity term, is responsible for the proximity of the reconstruction to the measurements. Therefore, we write it as $L_{\text{fidelity}}(E) = \left\|G(\vec{E}) - G(\vec{E}_{\text{true}})\right\|_2$, where $G(\vec{E})$ is the measurements vector of a field $\vec{E}$ and $\|\cdot\|_2$ is the $L_2$ norm. The second term keeps the field piecewise continuous, using total variation norm $\|\cdot\|_{TV}$, known to match this requirement[77]: $L_E(\vec{E}) = \left\|\vec{E}\right\|_{TV} = \left\|\nabla\vec{E}\right\|_1$ when $\|\cdot\|_1$ is the $L_1$ norm. The third term relates to the continuity of the magnetic field $\vec{H}$: $L_E(\vec{E}) = \frac{1}{i\omega\mu_0}\left\|\nabla\vec{H}\right\|_2 = \left\|\nabla(\vec{\nabla} \times \vec{E})\right\|_2$. The last term aims to zero the polarization density $\vec{P}$ outside the sample, assuming that we have the sample's structure (that we can measure and reconstruct in advance using standard electron tomography[44]): $L_P(\vec{E}) = \|I \cdot P\|_2 = \left\|I \cdot \left(\vec{\nabla} \times \vec{\nabla} \times \vec{E} - \left(\frac{\omega}{c}\right)^2 \vec{E}\right)\right\|_2$, where $I$ is 1 outside the sample and 0 inside it. The parameters $c_0, c_1, c_2, c_3$ are called hyper-parameters and need to be

chosen manually or using advanced methods[78]. The minimum loss can be found using the ADAM[79] method implemented in the PyTorch library[79].

The optical near fields used as references were computed in COMSOL Multiphysics (frequency domain) on PML-terminated domains with tabulated material dispersion; the mesh was refined until the fields in the region of interest were stable. We exported the complex electric field on a Cartesian grid and generated synthetic electron images by rotating the 3D fields and applying the PINEM forward-projection operator in MATLAB, using the same sampling as the simulation grid. The angular coverage was $2\pi$ (360°) because the forward model is not 180°-symmetric. PINET reconstructions directly used the theorem in Eq. 2. Unless noted otherwise, parameters were kept consistent across examples to facilitate reproduction.

**Discussion and outlook**

While our approach provides unique capabilities for near-field imaging, it is not without its challenges and limitations. We saw that the electron cannot interact with field components whose phase-velocity is higher than the electron velocity as they do not satisfy phase-matching for any angle of incidence. Therefore, the electron is sensitive only to a high-pass filtered field, and this is all that can be reconstructed (Fig. 1) without additional assumptions (Fig. 5). Nevertheless, as we saw above (Figs. 2-5) the high-pass filter maintains the small features and gives high-quality understanding of the full near-field. Additional important challenges arise from technical considerations such as limited rotation range and stability that affect the reconstructed field (Figs. 2,5). These limitations could be overcome using algorithmic improvements like the physics-inspired gradient descent optimization we suggested (Fig. 5) or additional assumptions about the near-field as in compressed sensing[80] and machine learning[81].

To implement PINET in current facilities, three special features need to be added to existing ultrafast TEMs (Fig. 3). The first is an interferometric setup used to extract the phase of every projection, as with the photonic free-electron modulator (PELM)[38–40]. The second is a tomography sample holder allowing a broad range of rotations of the sample (e.g., as in ref [50]). The third, which has not been implemented yet, is the ability to maintain the laser-sample configuration while rotating. This challenging condition may be achieved by a special cavity that will keep the optical mode interacting with the sample. Another option is to rotate the sample only along the laser axis while adjusting its polarization with a linear or circular polarizer.

Beyond the ability to sense and reconstruct the field inside the structure, PINET provides additional complementary capabilities to the leading near-field imaging techniques such as sSNOM[14] and PEEM[13,19]. Unlike sSNOM, which relies on scanning tips that can potentially alter the near-field being measured, PINET employs electrons that do not alter the electromagnetic near-fields while measuring them (as long as the fields are far from the quantum limit). Relative to PEEM, which is less efficient for low-energy photons, PINET is more sensitive at the lower energies, making it particularly promising in the mid-IR range, suitable for investigating phenomena like phonon polaritons in 2D materials. Generally, PINET is relevant to near-fields across a broad spectrum.

This work developed the concept of tomography for time-harmonic near-fields. However, many important electromagnetic effects cover a wide bandwidth, as with extreme non-linear optical effects like high-harmonic generation[82] and with ultrafast THz pulse generation based on optical rectification[83] or the photo-Dember effect[84]. These phenomena can even create single-cycle dynamics. In these cases, the near-fields interact with electrons in a more complicated way[85], which will require generalizing the PINET theory. To reveal the full dynamics of broad bandwidth or few cycle near-fields, a generalized 4D near-field tomography must be developed.